# A photonic integrated processor for multiple parallel computational tasks

Sheng Dong[1,2#], Ruiqi Zheng[1,2#], Huan Rao[1,2#], Junyi Zhang[1,2], Jingxu Chen[1,2], Chencheng Zeng[1,2], Yu Huang[1,2], Jiejun Zhang[1,2*] and Jianping Yao[1,2,3]

[1]Guangdong Provincial Key Laboratory of Optical Fiber Sensing and Communication, Institute of Photonics Technology, Jinan University, Guangzhou 510632, China
[2]College of Physics & Optoelectronic Engineering, Jinan University, Guangzhou 510632, China
[3]Microwave Photonics Research Laboratory, School of Electrical Engineering and Computer Science, University of Ottawa, Ottawa, ON K1N 6N5, Canada
email: zhangjiejun@jnu.edu.cn

**ABSTRACT**

Optical networks with parallel processing capabilities are significant in advancing high-speed data computing and large-scale data processing by providing ultra-width computational bandwidth. In this paper, we present a photonic integrated processor that can be segmented into multiple functional blocks, to enable compact and reconfigurable matrix operations for multiple parallel computational tasks. Fabricated on a silicon-on-insulator (SOI) platform, the photonic integrated processor supports fully reconfigurable optical matrix operations. By segmenting the chip into multiple functional blocks, it enables optical matrix operations of various sizes, offering great flexibility and scalability for parallel computational tasks. Specifically, we utilize this processor to perform optical convolution operations with various kernel sizes, including reconfigurable three-channel 1×1 convolution kernels and 2×2 real-valued convolution kernels, implemented within distinct segmented blocks of the chip. The multichannel optical 1×1 convolution operation is experimentally validated by using the deep residual U-Net, demonstrating precise segmentation of pneumonia lesion region in lung CT images. In addition, the capability of the 2×2 optical convolution operation is also experimentally validated by constructing an optical convolution layer and integrating an electrical fully connected layer, achieving ten-class classification of handwritten digit images. The photonic integrated processor features high scalability and robust parallel computational capability, positioning it a promising candidate for applications in optical neural networks.

## 1. INTRODUCTION

The rapid advancements in artificial intelligence, particularly in the field of deep learning, have significantly increased the demand for high-speed data computing and large-scale digital processing[1,2]. Convolutional neural networks (CNNs), a fundamental architecture in deep learning, are known for their capability to capture spatial dependencies and invariant features in diverse input data[3,4]. CNNs have demonstrated state-of-the-art performances in computer vision tasks, including image detection, object classification, and semantic segmentation[5-9]. The core of CNNs is the convolution operation, which relies on matrix multiplications between input tensors and sets of $n \times n$ kernels to extract feature maps. Improving the speed and scalability of these matrix computations is essential for advancing CNNs performance. However, conventional electronic computing systems based on the von Neumann architecture face inherent limitations due to the separation of memory and processing units. This design constrains hinders computational speed and increases power consumption, resulting in a bottleneck that restricts both processing performance and energy efficiency[10,11]. This architectural constraint poses significant challenges, particularly for data-intensive applications such as medical image analysis, autonomous driving, and large-scale object recognition[12].

Optical neural networks (ONNs) have gained attention as a promising alternative for next-generation neuromorphic hardware processors[13-15]. ONNs enable simultaneous optical signal transmission and image data processing, overcoming the bottlenecks of the von Neumann computing paradigm. By leveraging free-space and optical waveguide interconnections, ONNs can support parallel information processing with large operation bandwidth and ultra-high speed[16,17]. Furthermore, their ability to exploit the large bandwidth of photonic systems makes them particularly well-suited for accelerating large-scale matrix operations required in CNNs, which demand high-speed data transmission and large-scale parallel processing capability[18-20]. In recent years, ONNs have demonstrated their advantages of high energy efficiency, large bandwidth and high parallelism through techniques such as optical diffraction[21,22], optical interference[16,23] and time-wavelength multiplexing[19,24,25]. However, the implementation of the proposed systems relies mainly on discrete devices, which are space-intensive and pose challenges for large-scale implementation. While photonic integrated chip technologies offer improved matrix operation scalability and enhanced computing speed, they can often handle fixed, single-task computational tasks rather than supporting simultaneous parallel optical matrix operations at varying sizes. As the application scenarios become more complex, the architectures of CNNs are becoming increasingly massive. For example, the U-Net architecture, a special CNN, includes successive upsampling layers to achieve high-resolution outputs, while the ResNet architecture leverages residual connections to facilitate the training of deep networks. As one of the key components of versatile CNNs architectures, the convolution kernels have various sizes. For instance, 1×1 convolution kernels that convolves the input tensor with a series of filters corresponding to a matrix size of 1×1, typically used to allow information interaction between channels. On the other hand, 2×2 convolution kernels are used to reduce spatial dimensions while capturing local features. Therefore, there is an increasing need for computational systems that are both versatile and scalable to accommodate these demands. A photonic integrated processor that harnesses the unique properties of optical waveguides - such as wide bandwidth, low loss, small latency, and highly parallel processing capabilities - can support multiple parallel computational tasks, including simultaneous optical matrix operations with various sizes. This makes it ideal for a wide range of complicated applications such as optical computing, artificial intelligence, and real-time processing of communication signals. It also positions the technology for significant adoption in future artificial intelligence platforms that require more massive and intricate architecture.

In this paper, we present and demonstrate a photonic integrated processor designed for executing multiple parallel computational tasks, with a specific focus on its application in ONNs to facilitate compact and reconfigurable optical matrix operations. The photonic integrated processor is fabricated on a silicon-on-insulator (SOI) platform, which contains sixteen optical interference units, including Mach-Zehnder interference (MZI) and multimode interference (MMI) cells. The chip supports fully reconfigurable matrix operations by dynamically adjusting the phase shifters (PSs) on the MZIs. The processor can also be divided into multiple functional blocks, allowing for simultaneous execution of optical matrix operations of varying sizes. We utilize the photonic integrated processor to perform optical three-channel 1×1 optical convolution operations and 2×2 real-valued optical convolution operations using the time-wavelength multiplexing technique. For example, we demonstrate its capability by applying the 3-channel 1×1 convolution operation in a deep residual U-net to achieve precise segmentation of pneumonia lesion region in lung computed tomography (CT) images. Our results show that the deep residual U-net achieves a Dice coefficient of 0.658. Additionally, the photonic integrated processor is used to form an optical convolutional layer with two 2×2 convolution kernels, integrated with an electrical fully connected layer to form a photonic CNN. This configuration is experimentally validated through ten-class classification of handwritten digit images from the Modified National Institute of Standards and Technology (MNIST) dataset with an accuracy of 91.75%.

## 2.PRINCIPLE

The photonic integrated processor features a network architecture consisting of two-dimensional coupled optical switches. In contrast to conventional one-dimensional designs, where optical propagation is limited to a single axis, the two-dimensional configuration allows for denser integration of MZIs and MMIs in both horizontal and vertical directions. This design significantly enhances spatial efficiency, reducing the chip footprint while maintaining high computational performance and processing capacity. Fig. 1(a) illustrates the schematic diagram of the designed photonic integrated processor, featuring an arrayed optical MZIs supporting sixteen optical inputs and outputs. The sixteen input ports are connected to sixteen grating couplers (GCs), while eight of the sixteen output ports are connected to eight GCs and the other eight are connected to eight on-chip photodetectors (PDs). The photonic integrated processor consists of sixteen optical interference units, each comprising two MZI units and one crossing unit, as shown in Fig. 1(b). Fig. 1(c) and Fig. 1(d) show an MZI unit and a crossing unit. In total, the chip has thirty-two MZI units and sixteen crossing units. Each MZI unit consists of two 2×2 MMI couplers, two SOI strip waveguides with a PS in the upper waveguide, which is tunable based on thermo-optic effect using a heater, and two optical input and output ports. By adjusting the voltage applied to the heater, the refractive index of the PS is changed, creating a phase difference between the two arms of the MZI and modifying the coupling ratio at the output ports. The crossing units and the MZI units are interconnected via SOI strip waveguides and waveguide crossings, forming an intensive interconnected optical network. The large-scale architecture of this photonic integrated processor allows it to be segmented into multiple functional blocks for diverse computational tasks, particularly for reconfigurable matrix operations with different sizes.

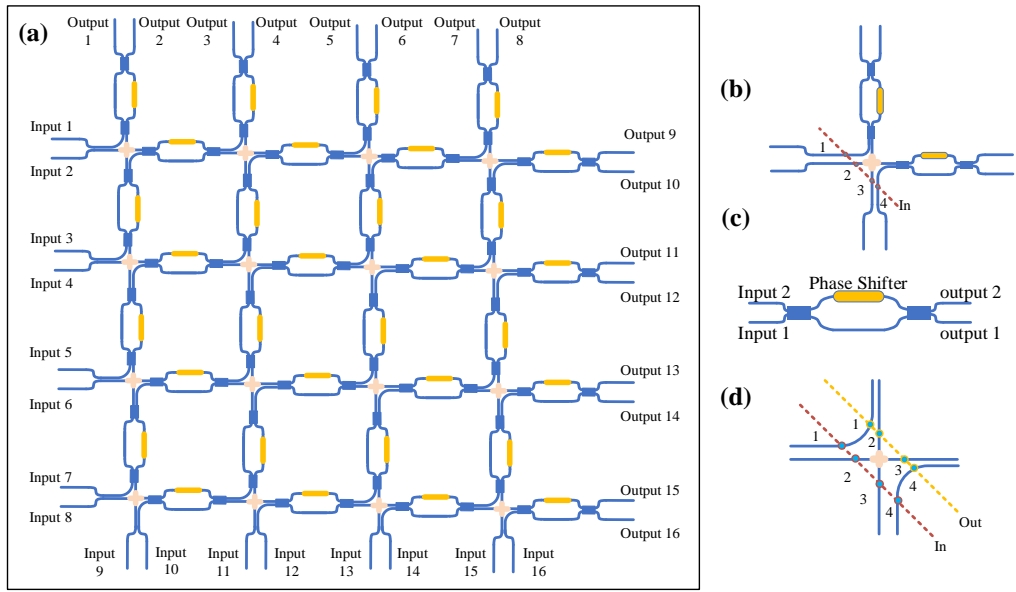

**Fig. 1 Schematic diagram of the photonic integrated processor. (a)** The schematic diagram of the photonic integrated processor; **(b)** an optical interference unit; **(c)** a Mach-Zehnder interferometer (MZI) unit; **(d)** a crossing unit.

In this photonic integrated processor, an MZI unit serves as a fundamental signal processing element. By controlling the optical power splitting ratio of the MZIs, specific transfer matrices can be realized, establishing a mapping relationship between the optical signals at sixteen input and sixteen output ports. Additionally, the chip can also be divided into multiple computational blocks, enabling parallel execution of different tasks with matrix operations of different sizes. Note that the values of the matrix elements can be tuned by changing the driving voltage applied to the thermo-optic PSs, resulting in dynamically reconfigurable transfer matrix operations (more details can be found in Supplementary Note 1). Consequently, the photonic integrated processor can execute multiple matrix operations, including multichannel 1×1 and 2×2 real-valued optical convolution kernels, significantly enhancing the processing speed and efficiency for large-scale datasets.

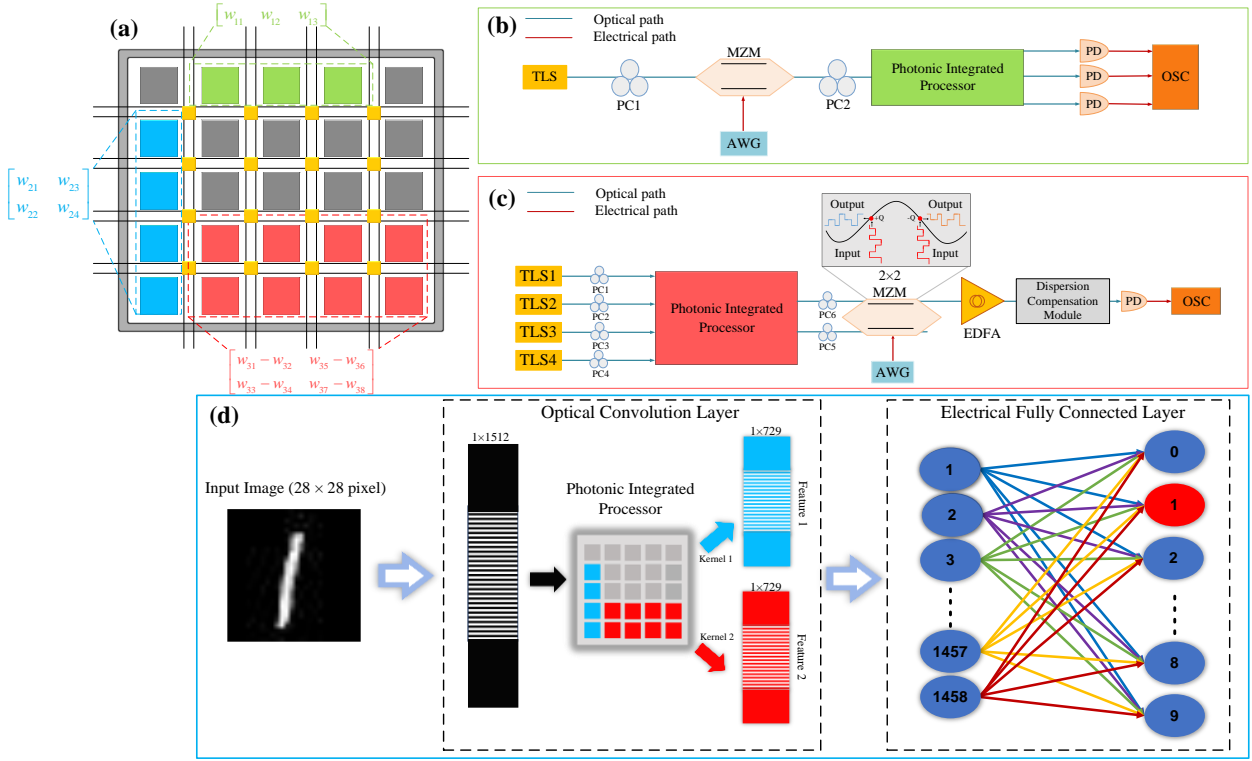

**Fig. 2 The photonic integrated processor that can be segmented into multiple functional blocks for parallel computational tasks.** (a) Different functional blocks represented by different colors for parallel tasks with varying matrix sizes, where the value of the matrix element $w_{m,n}$ is determined by considering the transfer matrix of the optical interference units. (b) Experimental setup for three-channel 1×1 optical convolution operations. (c) Experimental setup for 2×2 real-valued optical convolution operations. (d) The convolution neural network integrating an optical convolution layer and an electrical convolution layer for ten-class classification of handwritten digit images. TLS: tunable laser source PC: polarization controller, MZM: Mach-Zehnder modulator, AWG: arbitrary waveform generator, EDFA: erbium-doped fiber amplifier, OSC: oscilloscope, PD: photodetector.

To showcase the segmented, multi-functional and dynamically reconfigurable capabilities of the photonic integrated processor, we partition the chip into multiple functional blocks, each designed to perform a distinct optical computational task with a specific size. As shown in the Fig. 2(a), three functions are implemented through three active regions represented by three colors of green, red and blue, while the grey areas represent the inactive regions of the chip. The computational tasks include three-channel 1×1 convolution operations (green), 2×2 real-valued convolution operations for extracting the feature maps from digital images (red), and an optical convolutional layer with two 2×2 convolution kernels (red and blue). The latter is implemented by incorporating an electrical fully connected layer that performs the ReLU non-linear activation function to construct a CNN for ten-class classification of handwritten digit images ('0–9') from the MNIST dataset.

First, a three-channel 1×1 optical convolution operation using the green-colored functional part of the chip is studied. As shown in Fig. 2(b), a continuous-wave (CW) light is generated from a laser diode (LD) which is sent to a Mach-Zehnder modulator (MZM) via a polarization controller (PC1). Note that PC1 is used to minimize the polarization dependent loss. A two-dimensional (2D) image is flattened to be a one-dimensional (1D) vector, which is sent to an arbitrary waveform generator (AWG) to produce an electronic signal. This electronic signal is applied to the MZM to generate an intensity-modulated optical signal. The optical signal from the MZM is sent into the photonic integrated processor via a PC (PC2). Because the GC used for vertical fiber-to-chip coupling is sensitive to the polarization state of an optical signal into the chip, PC2 is employed before the GC to minimize the polarization dependent loss. After traveling through six passing MZI units, the amplitude of the signal is modified by the six passing MZI units, shown as the green-colored functional part in the processor. The PSs in these MZI units are adjusted according to the convolution weight values of three 1×1 kernels, achieving a 1×3 optical convolution kernel corresponding to a 1×3 matrix, as illustrated in the Fig. 2(a). The processed optical signals are sent out of the chip via three output ports, which are detected at three off-chip PDs to generate three electrical signals. The electrical signals are then sampled by an oscilloscope (OSC). Note that each output electronic 1D vector represents the result of the three-channel 1×1 optical convolution operation.

Then, a 2×2 real-valued optical convolution operation using the red-colored functional part of the chip is investigated. As demonstrated in Fig. 2(c), four wavelengths generated by an array of LDs are coupled into the processor via four GCs. Again, to minimize the polarization dependent loss at the GCs, a PC is incorporated between a laser source and a GC. Within the red-colored functional part, each wavelength is split into two paths. Dynamic voltage adjustments are applied to the PSs in the MZIs, to control the powers of the four wavelengths corresponding to the weights in the 2×2 kernel. The optical signals are then transmitted out of the chip via two output ports, with four wavelengths exiting from each port. These wavelengths are subsequently applied to a 2×2 MZM through two input ports. The modulation signal is obtained by flatting the 2D image to be a 1D vector. At the MZM, the four wavelengths are modulated by the electrical signal from an AWG. As illustrated in the insert of the Fig. 2(c), the 2×2 MZM are biased at the quadrature points with positive and negative slopes. By leveraging the reconfigurable matrix operation capability of the chip to achieve the power difference of the same wavelength signal at two output ports, matrix with both negative and positive elements are achieved (more details can be found in Supplementary Note 3). To compensate for the power loss along the optical transmission channel, an erbium-doped fiber amplifier (EDFA) is incorporated. Using the time–wavelength multiplexing technique, a dispersion compensation module is employed to introduce a time delay between two adjacent wavelengths which is equal to the reciprocal of the baud rate. This fixed time delay corresponds to a one-symbol offset between each optical signal. The time-delayed signals are detected at a PD and sampled by an OSC, where convolution operations between the flattened image and real-valued 2×2 kernels are performed (more details can be found in Supplementary Note 2).

Next, the photonic integrated processor is used in conjunction with an electrical fully connected layer in a digital computer to form a photonic CNN. As illustrated in Fig. 2(d), the blue-colored functional part of the chip is used to realize a 2×2 convolution kernel based on the experimental setup in the Fig. 2(c). Different from the red-colored region, this part utilizes four input ports and one output port of the chip, supporting convolution kernels with positive weights. The optical convolution layer, which contains two convolution kernels, is then combined with the electrical fully connected layer to implement a ten-class classification task of "0 ~ 9" handwritten digit images. The kernel with real-valued weights is implemented in the red-colored region and another kernel with positive-valued weights in the blue-colored region is utilized to extract the feature maps of a digital image and generate 1D feature vectors. The feature maps are activated by the ReLU nonlinear activation function and then fed into the fully connected layer for classification. The final output contains 10 values, with the largest value in the $1 \times 10$ vector indicating the predicted number.

## 3. EXPERIMENTAL RESULT

The photonic integrated processor is fabricated on a SOI platform. A microscope image of the processor is shown in the Fig. 3(a). The chip consists of thirty-two interconnected MZI units and sixteen crossing units, with a total of sixteen input ports and sixteen output ports. The input ports are connected to sixteen GCs, and eight of the output ports are connected to eight GCs and the other output ports are connected to eight on-chip PDs. The GCs utilize vertical-to-chip coupling with a spacing of 127 μm, and a solder pad has a size of 75 μm × 55 μm. Each MZI unit contains a PS, with a length of 100 μm, which is tunable based on the thermo-optic effect. An image of the packaged chip is shown in Fig. 3(b) (more details can be found in Supplementary Note 4).

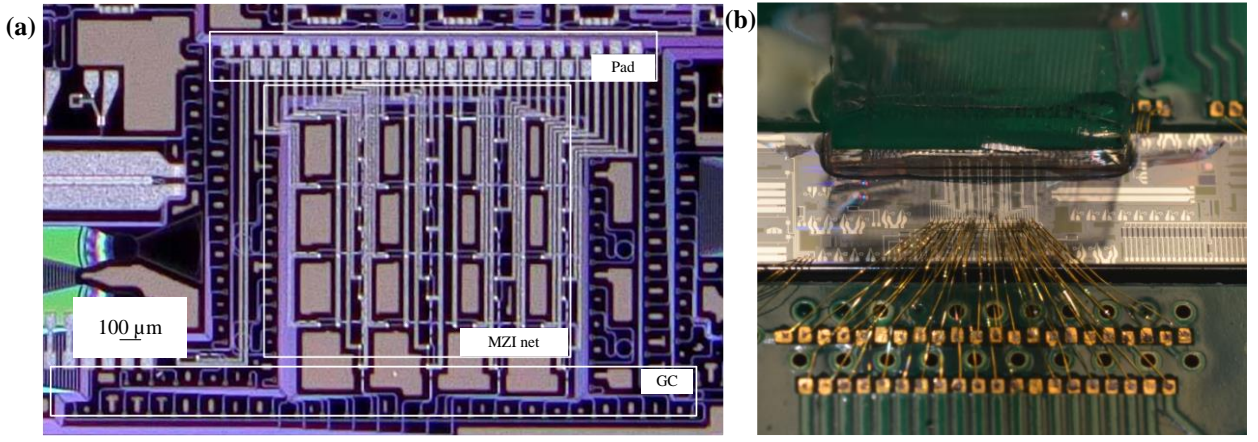

**Fig. 3 Images of the photonic integrated processor. (a)** The microscope image of the photonic integrated processor. **(b)** The image of the packaged chip. MZI: Mach-Zehnder interferometer, GC: grating coupler

In the experiment based on the setup shown in the Fig. 2(b), three-channel 1×1 convolution operation in a deep residual U-Net for a biomedical image segmentation task is implemented. Images from COVID-SemiSeg dataset are selected and divided into a training set and a test set with a ratio of 9:1. The processed data are 8-bit grayscale CT images with a resolution of 512×512, which is flattened into 1D vector and encoded to the AWG (Tektronix AWG70002A) at a data rate of 5 GBaud/s. The powers of the three optical wavelengths are adjusted according to the convolution weight values of the three 1×1 kernels. Detected by off-chip PDs and sampled by an OSC (Teledyne LeCroy LabMaster 10-36Zi), the collected data are utilized to train a deep residual U-net (more details can be found in Supplementary Note 5), and the training result is shown in Fig. 4. The model achieves a Dice coefficient of 0.658, a sensitivity (true positive rate) of 0.565, a specificity (true negative rate) of 0.868, and a mean absolute error (MAE) of 0.0733. These results demonstrate that the photonic integrated processor can process biomedical images with outstanding performance, meeting the requirements for practical applications.

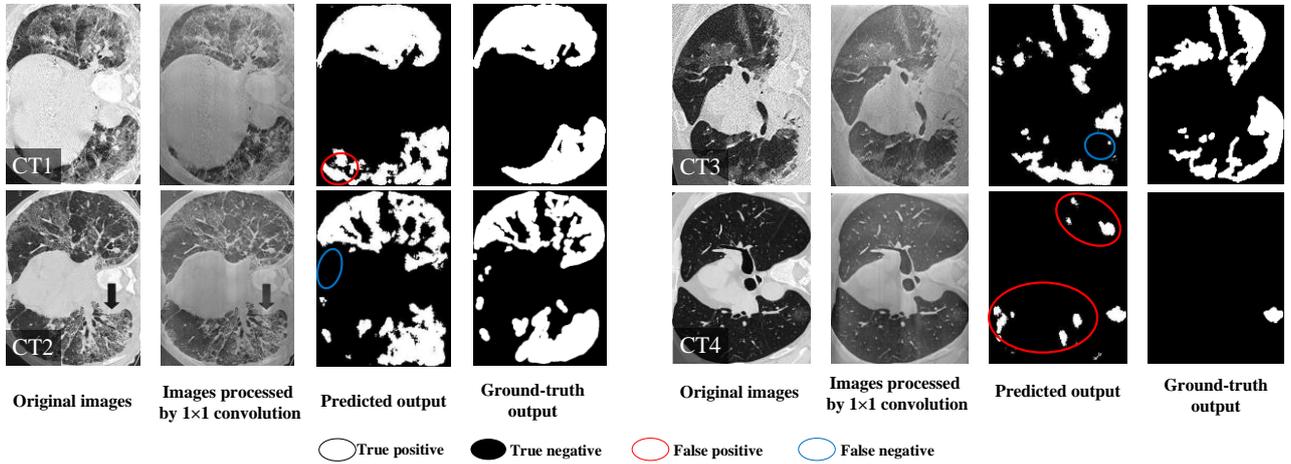

**Fig. 4** Comparison between the predicted results (partial) of the CT images and the ground-truth images.

To implement the 2×2 real-valued optical convolution operation using the photonic integrated processor, an experiment based on the experimental setup in the Fig. 2(c) is performed. Four optical wavelengths are generated from an array of LDs at 1544.996, 1546.416, 1547.832, and 1549.217 nm. Each wavelength is coupled into the chip through a PC to minimize the polarization dependent loss. The powers of the four wavelengths are dynamically controlled by adjusting the voltages applied to the PSs within the MZIs in the red-colored functional part in Fig. 2(a), enabling reconfigurable optical matrix operations. For matrix elements with both positive and negative values, the same wavelength signals at two output ports of different power are sent to a MZM via two input ports, to achieve real-valued positive and negative matrix operations. 2D digital images are flattened to 1D signal and encoded to the AWG (Keysight M8194A). The electronic signal from the AWG is applied to the 2×2 MZM at a data rate of 10 GBaud/s to modulate optical wavelengths. A dispersion compensation module with an average dispersion parameter of -70.68 ps/nm at 1545 nm is used to achieve an identical time delay difference, corresponding to one-symbol interval between adjacent signals. This fixed time delay allows the transformation of optical matrix operations into optical convolution when the modulated optical signals are detected at a PD [19]. The capability of the system for the computational tasks about 2×2 real-valued optical convolution operations is demonstrated with the results shown in the Fig. 5(a)-(c), where the original images of five images from the MNIST dataset ("2", "5", "7" and "9") are shown in Fig. 5(a), the feature maps obtained with a digital computer are shown in Fig. 5(b) and the feature maps from the photonic integrated processor are shown in Fig. 5(c). Comparing the feature maps obtained from the digital computer with those obtained by the photonic integrated processor, the latter presents an average root mean square error (RMSE) of 0.159 among the 12 feature images shown in Fig. 5.

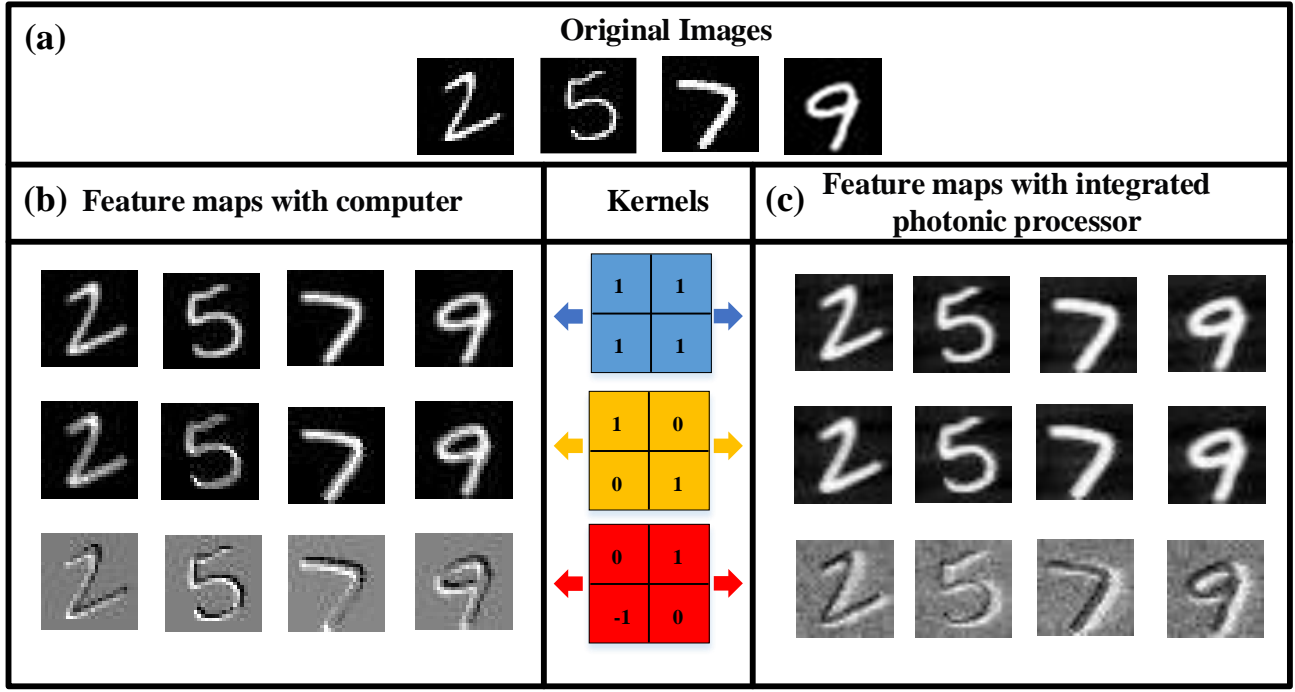

**Fig. 5 Feature images obtained when using 2×2 real-valued reconfigurable optical convolution kernels.** (a) The input handwritten digit images of "2", "5", "7" and "9". (b) Feature maps based on a digital computer. (c) Feature maps based on the photonic integrated processor.

Subsequently, the complete CNN shown in Fig. 2(d), is trained with 60,000 images and tested with 10,000 images from the MNIST dataset in a digital computer to implement ten-class classification of handwritten digit images ('0–9'). After completion of the training process, the structure and weight paraments of the neural network are fixed (more details can be found in Supplementary Note 6). Fig. 6(a) shows the variation of accuracy and cross-entropy loss corresponding to the increasing training epochs, and a maximum of 50 training epochs is applied. Next, a total of 470 handwritten digit images are tested in the photonic CNN, and the confusion matrix for the test images is shown in the Fig. 6(b). The classification results show that an accuracy of 91.75% is achieved for the experiment compared to 93.48% for the theoretical model. The deviation of 1.73% from the theoretical accuracy is mainly caused by two factors, the electrical and optical noise and the instability of the optical devices.

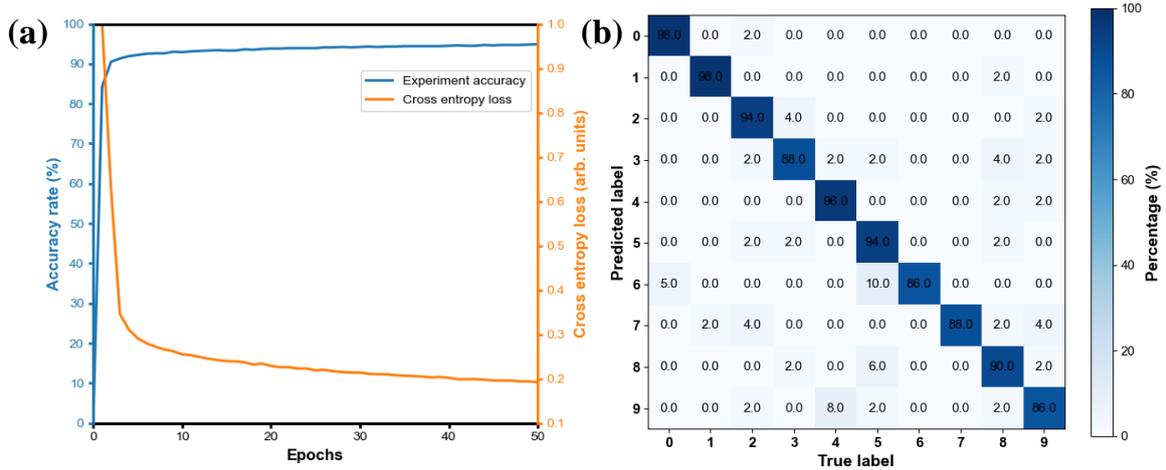

**Fig. 6 MNIST handwritten digit image classification.** (a) The variation of accuracy and cross entropy loss during 50 epochs of training. (b) The confusion matrix of recognizing 470 handwritten digit images through the photonic CNN, where the abscissa indicates the true labels and the ordinate indicates the recognition results.

## 4. CONCLUSION

We have demonstrated a photonic integrated processor capable of supporting multiple computational tasks by segmenting the chip into multifunctional blocks. Fabricated on a SOI platform, this processor has sixteen optical interference units based on thirty-two MZIs and sixty-four MMI cells, enabling dynamical and reconfigurable matrix operations. We used the photonic integrated processor to perform convolutions with reconfigurable three-channel 1×1 convolution kernels and 2×2 real-valued convolution kernels in the segmented parts of the chip. The multichannel optical 1×1 convolution operation was experimentally validated by using the deep residual U-Net to achieve precise segmentation of pneumonia lesion region in lung CT images. The capability of 2×2 optical convolution operation was also experimentally validated by forming an optical convolution layer incorporating an electrical fully connected layer to achieve ten-class classification of handwritten digit images. The segmented, multifunctional, and parallel processing capabilities of the processor allows high-speed processing of signals and images, significantly reducing the computational burden on a hybrid computing system where the convolution operations can be implemented optically at high speed while other operations can be implemented electrically. This work contributes to the development of high-performance, scalable artificial intelligence networks capable of supporting complex and massive computations.

## ACKNOWLEDGEMENT

This work was supported in part by the National Key Research and Development Program of China under Grant 2021YFB2800804, in part by the Guangdong Province Key Field Research and Development Program under Project 2020B0101110002, and in part by the Guangdong Engineering Technology Research Center for Integrated Space-Terrestrial Wireless Optical Communication. The authors thank the Advanced Micro Foundry, Singapore for PIC fabrication.

Supporting Information for
# A photonic integrated processor for multiple parallel computational tasks

Sheng Dong[1,2#], Ruiqi Zheng[1,2#], Huan Rao[1,2#], Junyi Zhang[1,2], Jingxu Chen1,2, Chencheng Zeng[1,2], Yu Huang[1,2], Jiejun Zhang[1,2*] and Jianping Yao[1,2,3]

# A photonic integrated processor for multiple parallel computational tasks

**Authors:**
Sheng Dong[1,2,#], Ruiqi Zheng[1,2,#], Huan Rao[1,2,#], Junyi Zhang[1,2], Jingxu Chen[1,2], Chencheng Zeng[1,2], Yu Huang[1,2], Jiejun Zhang[1,2,*] and Jianping Yao[1,2,3]

**Affiliations:**
[1]Guangdong Provincial Key Laboratory of Optical Fiber Sensing and Communication, Institute of Photonics Technology, Jinan University, Guangzhou 510632, China
[2]College of Physics & Optoelectronic Engineering, Jinan University, Guangzhou 510632, China
[3]Microwave Photonics Research Laboratory, School of Electrical Engineering and Computer Science, University of Ottawa, Ottawa, ON K1N 6N5, Canada

**E-mail:**
email: zhangjiejun@jnu.edu.cn

# Table of Contents

# Supplementary Information

**Supplementary Note 1: Implementation of optical convolution kernels**

In this optical photonic integrated processor, each MZI unit serves as a fundamental signal processing element. By precisely controlling the split ratio of each MZI, specific transfer matrices can be realized. The values of the matrix elements are determined by the transfer matrix of the optical interference units, the MZI units and their connection architecture. The two-dimensional configuration allows for denser integration of MZIs and MMIs in both horizontal and vertical directions. The chip is allowed to perform reconfigurable optical convolution operations with different sizes in the segmented parts.

For example, the computational task of a three-channel optical convolution operation can be mathematically represented as

$$\begin{bmatrix} O_1 \\ O_2 \\ O_3 \end{bmatrix} = \begin{bmatrix} \omega_{01} \\ \omega_{02} \\ \omega_{03} \end{bmatrix} \cdot I_0 \tag{1}$$

where the $I_0$ and the $O_n$ (1≤$n$≤3) and matrix element $\omega_{0n}$ (1≤$n$≤3) represent the input and output optical signals, and three reconfigurable matrix elements determined by the convolution kernels can be controlled by the integrated photonic processor.

Then, the computational task of a 2×2 optical convolution operation can be mathematically represented as

$$O_4 = \begin{bmatrix} \omega_1 & \omega_2 & \omega_3 & \omega_4 \end{bmatrix} \cdot \begin{pmatrix} I_1 \\ I_2 \\ I_3 \\ I_4 \end{pmatrix} \tag{2}$$

where the $I_n$ (1≤$n$≤4) and the $O_4$ represent the input and output optical signals, and matrix elements $\omega_n$ (1≤$n$≤4) indicate the response of the processor and only positive-valued elements can be achieved. By biasing the 2×2 MZM at the quadrature points with positive and negative slopes, the optical matrix operations with negative and positive elements are enabled. The 2×2 real-valued matrix is given by

$$O_5 = \begin{bmatrix} \omega_{51} - \omega_{52} & \omega_{61} - \omega_{62} & \omega_{71} - \omega_{72} & \omega_{81} - \omega_{82} \end{bmatrix} \cdot \begin{pmatrix} I_5 \\ I_6 \\ I_7 \\ I_8 \end{pmatrix} \tag{3}$$

where the $I_n$ (5≤$n$≤8) and the $O_5$ represent the input and output optical signals, and the matrix elements $\omega_{m,n}$ (5≤ $m$ ≤8, 1≤$n$≤2) indicate the response of the processor, where both positive and negative value elements can be achieved.

**Supplementary Note 2: Image flattening and convolution data processing**

The 2D image flattening and significant convolution result extraction can be represented with formulas. For an input image with $N \times M$ pixels, by configuring the core size of the photonic integrated processor to 2×2, the input image can be divided into $(N-(2-1)) = (N-1)$ sub-images by rows, with each sub-image containing two rows of the input image (for example, each sub-image has 2×$K$ pixels). Then, each sub-image is flattened into a $1 \times 2M$ vector by columns, and ($N-1$) sub-images are concatenated head-to-tail to form a $1 \times (M \times (N-1))$ vector. At the output from the photonic integrated processor, convolution results are obtained. Note that the coordinates of the significant values in the convolution results can be expressed as

$$L = 2i + 2m(n-1), (i = 1, 2, 3..., M; n = 1, 2, 3, ..., N-1) \tag{4}$$

Taking Supplementary Fig. 1 as an example, a 4×4 image is convolved with a 2×2 kernel in one channel. First, the input image is divided into 3 sub-images, each containing two adjacent rows, and then these sub-images are flattened into three 1×8 vectors by columns. Subsequently, these three 1×8 vectors are concatenated head-to-tail to form a 1×24 vector to complete the convolution operation. According to Equation (1), $N = M = 4$, and the significant values are $[y_4 \; y_6 \; y_8 \; y_{12} \; y_{14} \; y_{16} \; y_{20} \; y_{22} \; y_{24}]$ (marked with a cyan background in Supplementary Fig. 1). The feature map can be obtained by reshaping these significant values into a 3×3 matrix given by

$$\begin{bmatrix} y_4 & y_6 & y_8 \\ y_{12} & y_{14} & y_{16} \\ y_{20} & y_{22} & y_{24} \end{bmatrix} \tag{5}$$

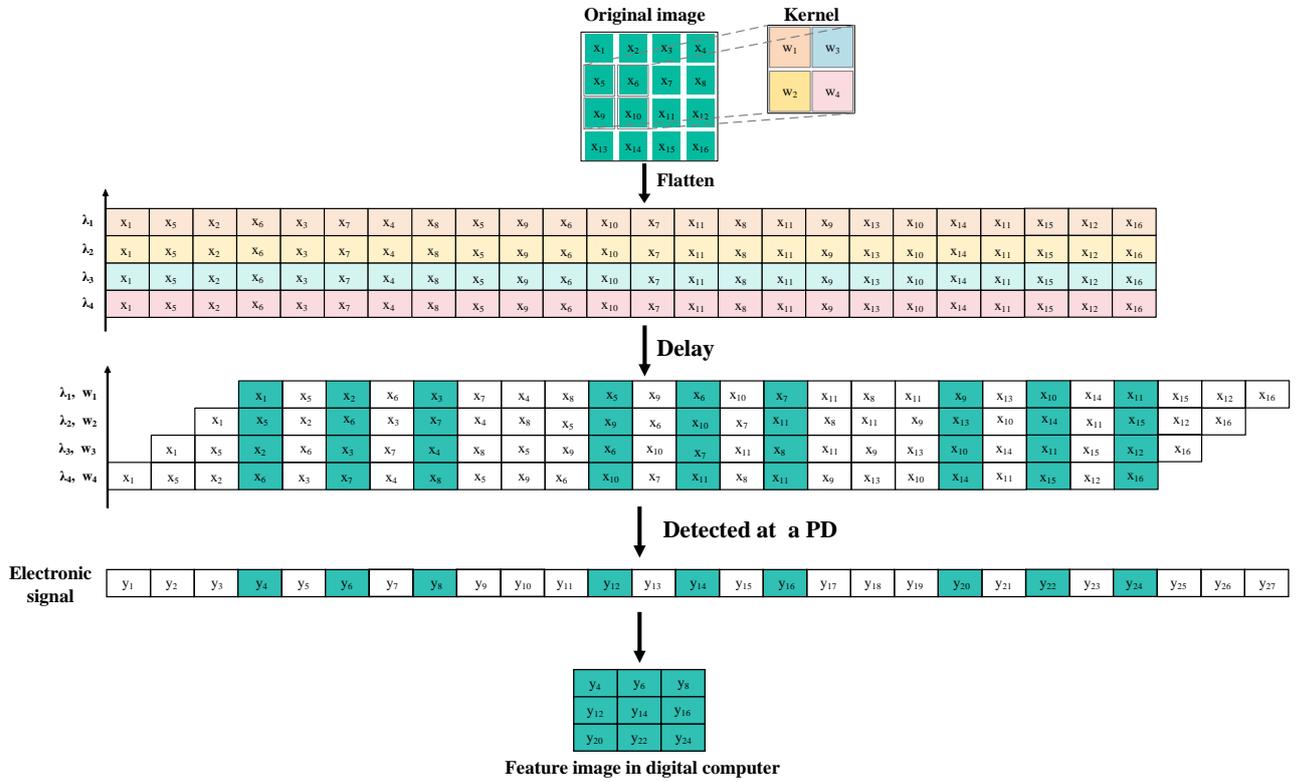

Supplementary Fig. 1 Optical convolution result extraction.

## Supplementary Note 3: Propagation and control of optical signals within the processor

The photonic integrated processor has sixteen input ports and sixteen output ports, and sixteen optical interference units, which allow it to support multiple computational tasks. In the experimental setup, we utilize three function regions to perform optical matrix operations. For region 1, as shown in the Supplementary Fig. 2(a), the green-colored part, an optical signal carried by one wavelength is sent to the processor, which is split into three paths to achieve intensity modulation where the weights are applied based on the 1×3 convolution kernel and the results are obtained at the three output ports. For region 2, as shown in the Supplementary Fig. 2(b), the blue-colored part, an optical signal carried by four wavelengths are sent the processor, with each passing an MZI, to obtain the convolution results based on the 2×2 convolution kernel. Note that optical convolution operations with only positive values are achieved. For region 3, as shown in the Supplementary Fig. 2(c), the red-colored part, an optical signal carried by four wavelengths are sent to the processor. It is different from the blue region, each wavelength is split into two paths, achieving intensity modulation. The optical signals are sent out to two output ports, each of which contains four wavelengths. By incorporating a 2×2 MZM biased at the quadrature points with positive and negative slopes, optical convolution operations with both negative and positive values are achieved.

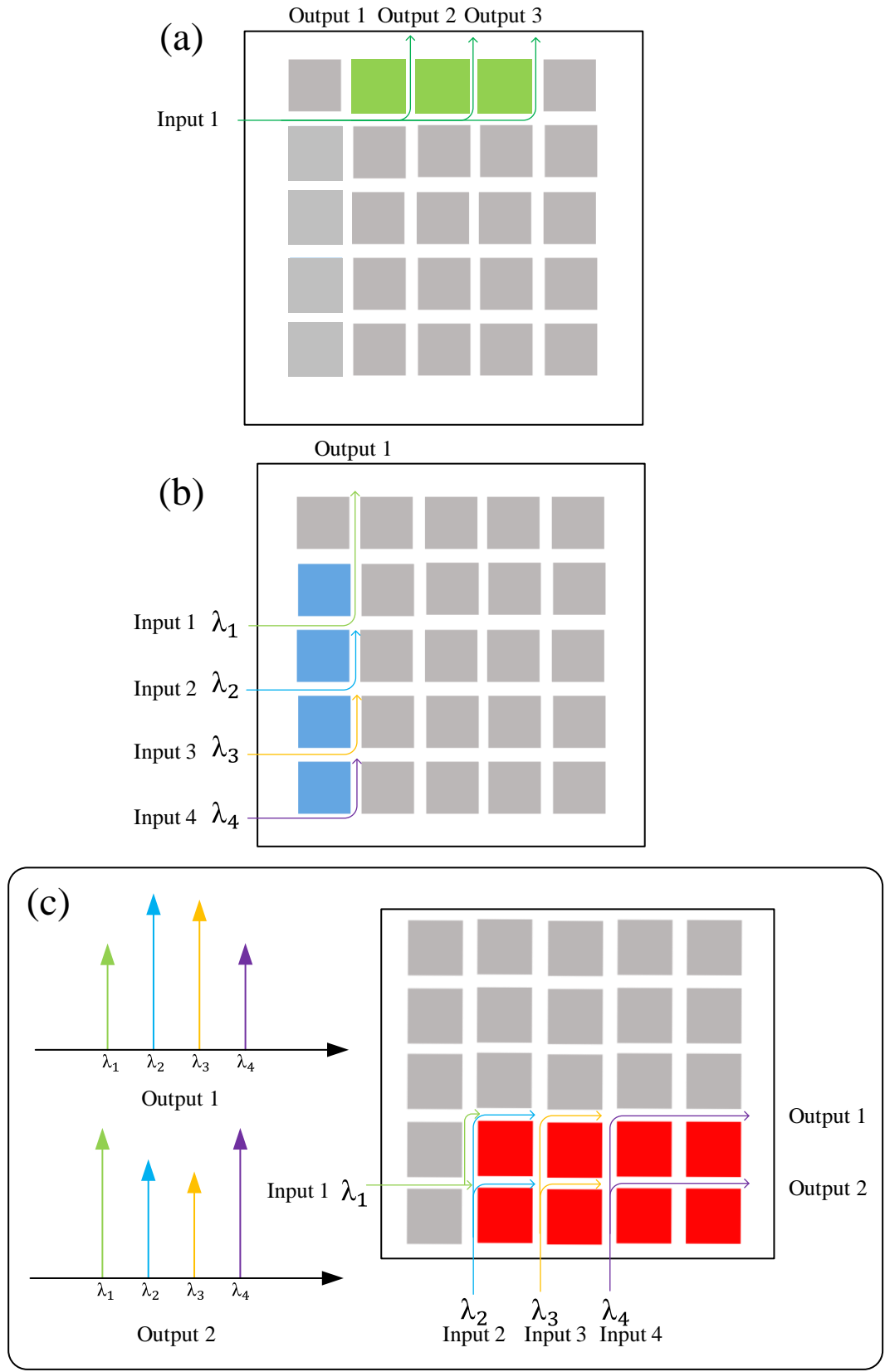

Supplementary Fig. 2 The propagation and control of optical signals within the photonic integrated processor.

**Supplementary Note 4：Chip packaging**

The layout design of the photonic integrated processor is shown in the Supplementary Fig. 3. Adjacent grating couplers (GCs) in the GC array is spaced at $127\mu m$, ensuring alignment between the optical fiber array and the GCs array. To meet the packaging requirements, the GCs are neatly arranged on one side, maintaining a distance of at least 1 mm away from the solder pads used for the gold wire bonding process, thereby enabling the GCs to input and output optical signals through the same optical fiber array. Considering that the photonic integrated processor contains thirty-two MZIs with each having two electrodes, using the traditional probe-based method not only leads to poor contact but also causes irreversible damage to the chip due to the repeated probe pressing. An electrical packaging solution is adopted, by which all electrodes on the photonic integrated processor are connected to the external printed circuit board (PCB) via gold wire bonding technology. The PCB is made of FR4 material, and the gold wire has a diameter of $25\ \mu m$. This packaging approach enhances the usability of the chip and improves its reliability for practical applications.

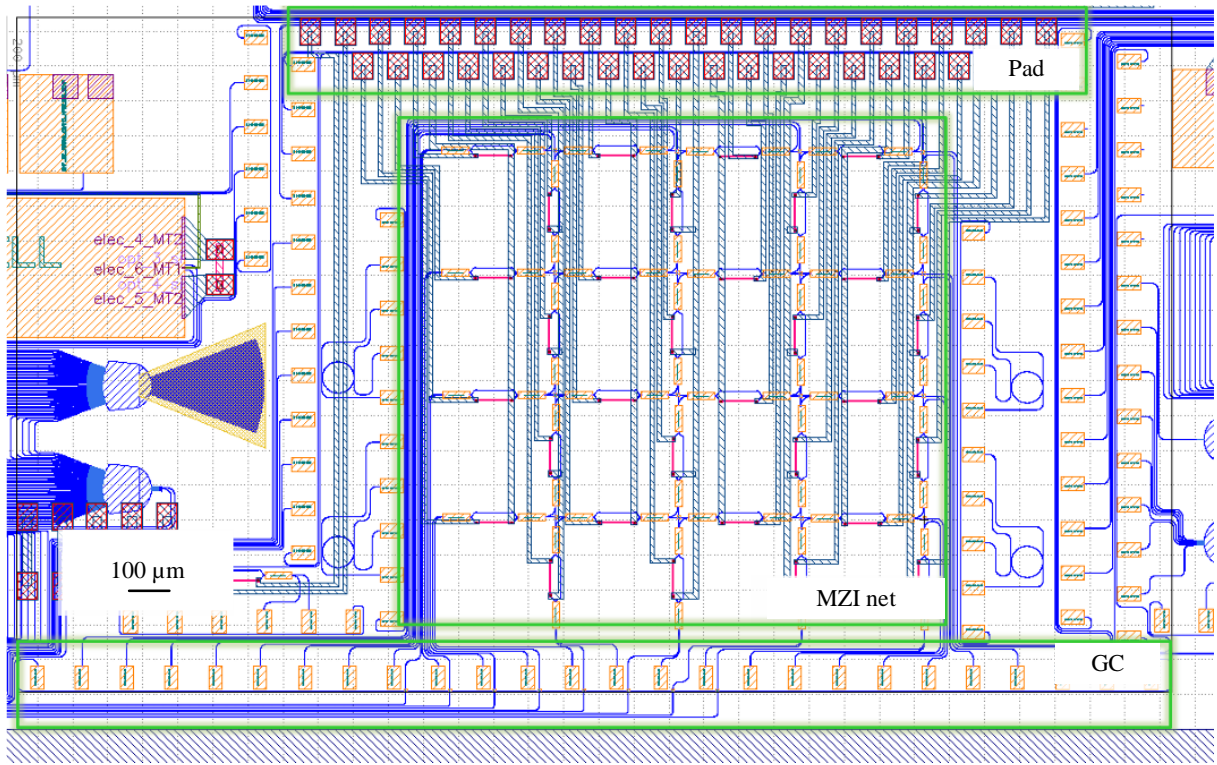

Supplementary Fig. 3 The layout design of the photonic integrated processor.

**Supplementary Note 5: U-Net network training**
- **Network architecture design**

The U-Net network, a popular deep learning architecture utilized for precise segmentation of COVID-19 lesion areas in lung CT images, as depicted in Supplementary Fig. 4, comprises 4 downsampling operations, 4 upsampling operations, and 4 skip connections.

Downsampling: In the decoder part of the network, starting from layer L1, the number of channels gradually increases, with 3×3×3 convolutions and downsampling operations with a stride of 2 to retain information; from L1 to L5, the number of channels increases to 1024, and each layer extracts the features through convolutions.

Upsampling: In the decoder part of the network, when the number of channels reaches 1024, the size of the feature maps is at its minimum, and then the resolution is restored through upsampling; in the final stage of upsampling, the number of channels is reduced to 64, and the output is converted into probability values through 1×1 convolutions and a sigmoid layer.

Skip Connections: At each upsampling stage in the decoder, connections are made with the corresponding stage's feature maps from the encoder. These skip connections fuse high-level and low-level features through concatenation, thereby improving the accuracy of segmentation.

- **Network training settings**

Data Preprocessing: Utilizing the pulmonary infection dataset from the COVID-SemiSeg Dataset, the data is randomly divided into training and testing sets with a ratio of 9:1.

Learning Rate: The Adam optimizer is used with an initial learning rate of 0.001.

Batch Size: It is set to 50 to balance memory usage and training efficiency.

Number of Training Epochs: A total of 300 training epochs.

- **Training platform**

Hardware Platform: Nvidia RTX4070ti GPU, Intel i5-13600KF is used, providing powerful parallel computing capabilities.

Software Platform: Based on the PyTorch framework, which simplifies model building and training processes with its high-level APIs.

The deep residual U-net running on the photonic integrated processor achieve the following performance metrics, a Dice coefficient of 0.658, a sensitivity (true positive rate) of 0.565, a specificity (true negative rate) of 0.868, and a mean absolute error of 0.0733.

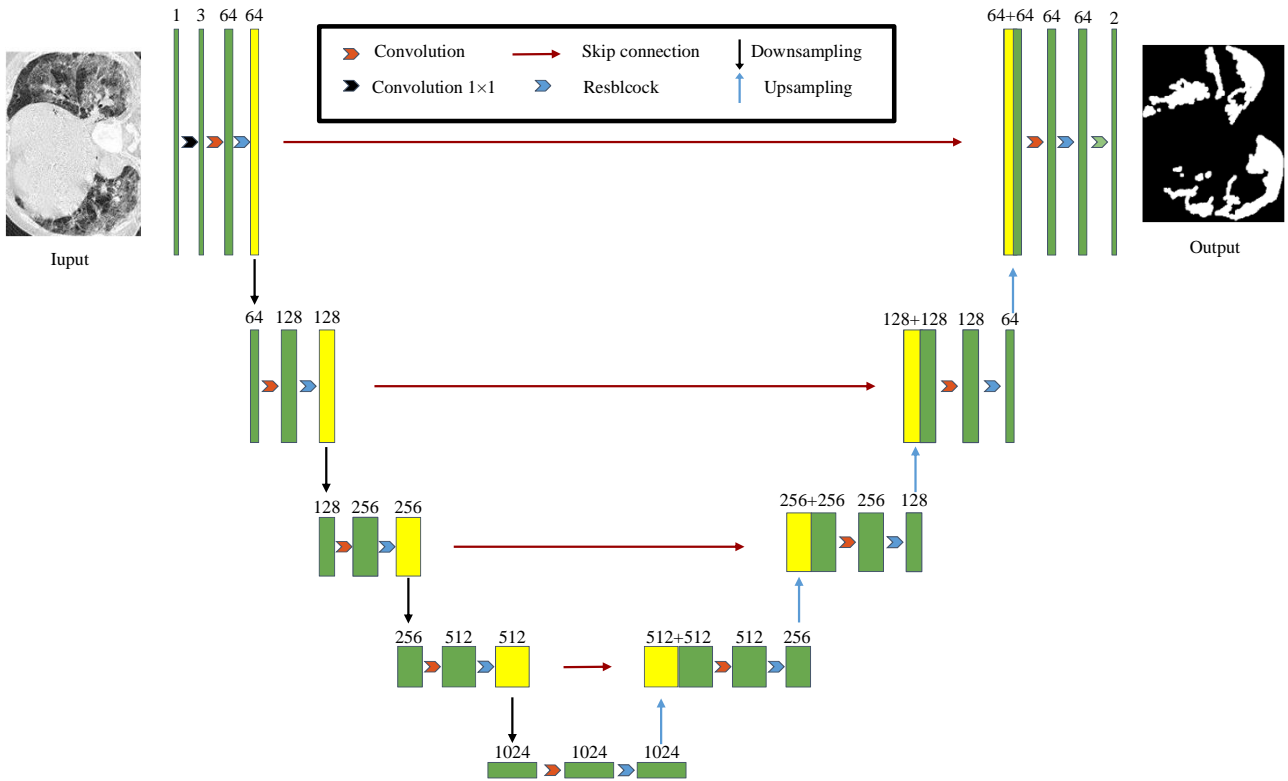

Supplementary Fig. 4 The architecture of the U-net.

**Supplementary Note 6: CNN training**

- **Network architecture design**

The convolutional neural network (CNN) model consists of multiple layers to accommodate the task of handwritten digit image recognition, as shown in Supplementary Fig. 5. The network structure is as follows:

Input Layer: Receives grayscale images with a size of 28×28×1.

Convolutional Layer: Contains 2 convolutional kernels of size 2×2, with a stride of 1, no edge padding, and no bias.

Activation Layer: Uses the ReLU activation function.

Fully Connected Layer: With an input feature number of 1458, it outputs 10 class classifications, corresponding to 0-9.

- **Network training settings**

Data Preprocessing: The MNIST dataset is divided into training and testing sets with a ratio of 8:2. The training set is used for model training, while the testing set is used for validation.

Learning Rate: Adam optimizer is used with an initial learning rate of 0.001.

Batch Size: It is set to 300 to balance memory usage and training efficiency.

Number of Epochs: A total of 50 training cycles.

Loss Function: Since it is a multi-class classification problem, we use the cross-entropy loss function.

- **Training platform**

Hardware Platform: Nvidia RTX4070ti GPU, Intel i5-13600KF is used, providing powerful parallel computing capabilities.

Software Platform: Based on the PyTorch framework, which simplifies model building and training processes with its high-level APIs (application programming interfaces).

The average accuracy of the CNN in recognizing the 10,000 handwritten digit images in the MNIST validation set is 0.9359, with detailed recognition rates for each category shown in Supplementary Fig. 6.

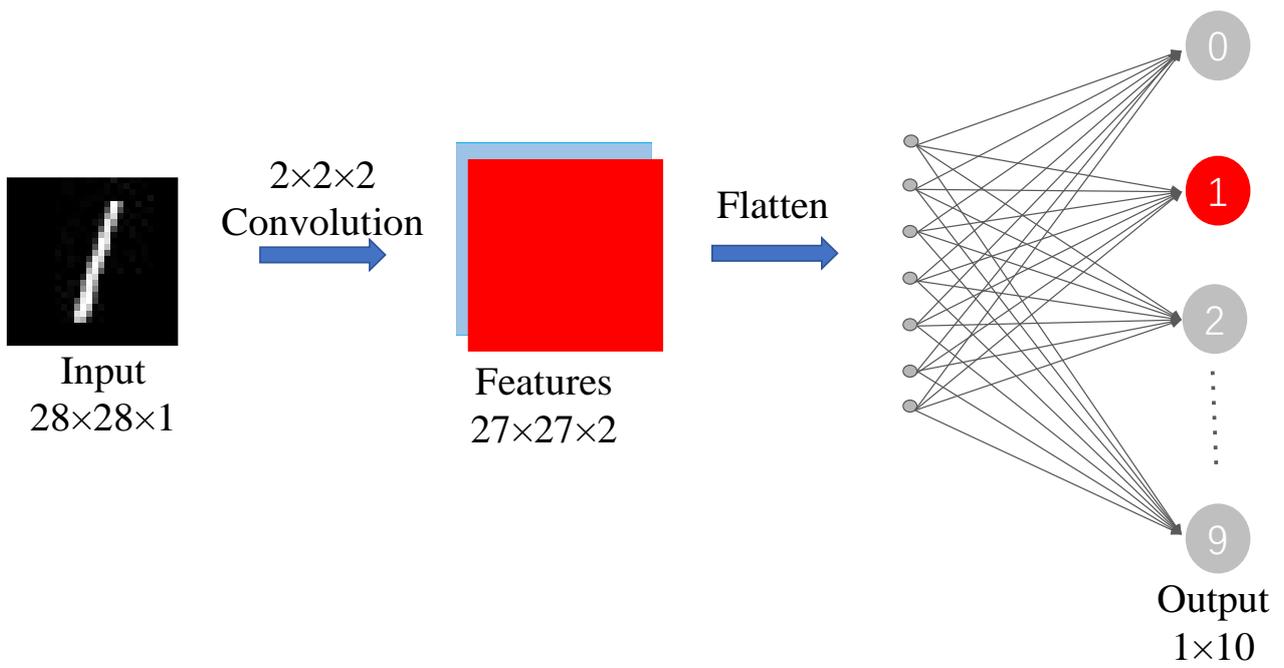

Supplementary Fig. 5 The architecture of the CNN.

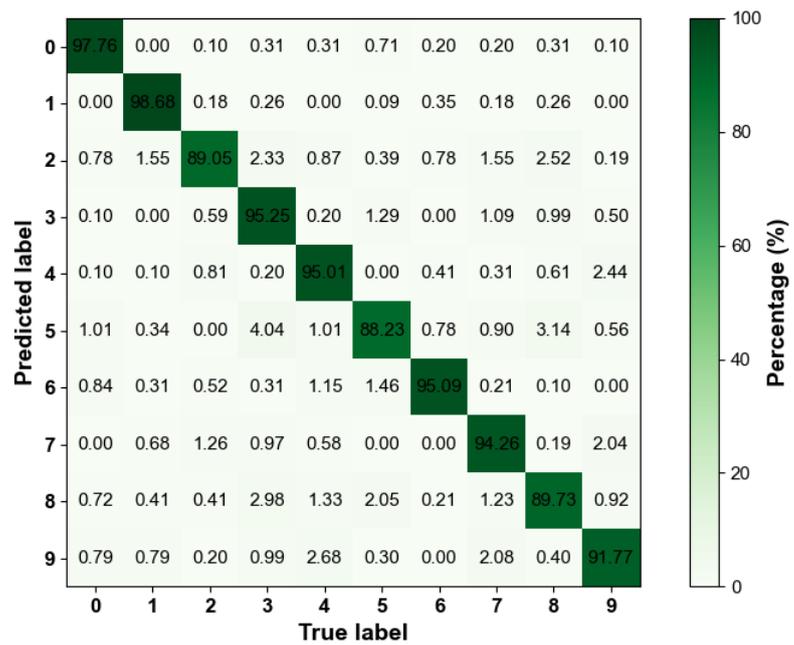

Supplementary Fig. 6 Recognition rates for each category in the MNIST validation set.